\documentclass[12pt]{amsart}
\usepackage{color}
\usepackage{amssymb}
\usepackage{hyperref}
\usepackage{dsfont,soul}
\usepackage{enumitem}
\pdfoutput=1
\setlist[enumerate]{labelsep=*, leftmargin=1.5pc,
topsep=1ex plus0.5ex minus0.2ex,
itemsep=1ex plus0.5ex minus0.2ex,
font=\rmfamily,
font=\upshape}
\newtheorem{thm}{Theorem}[section]
\newtheorem{cor}[thm]{Corollary}
\newtheorem{lem}[thm]{Lemma}
\newtheorem{pro}[thm]{Proposition}

\theoremstyle{definition}
\newtheorem{defn}[thm]{Definition}
\newtheorem{ex}[thm]{Example}
\newtheorem{fa}[thm]{Fact}
\newtheorem{rem}[thm]{Remark}
\numberwithin{equation}{section}
\newcommand{\id}{\mathds{1}}

\newcommand{\trace}{\operatorname{tr}}

\newcommand{\rank}{\operatorname{rank}}

\newcommand{\conv}{\operatorname{conv}}

\newcommand{\ii}{{\it i}}

\newcommand{\ri}{\operatorname{ri}}
\newcommand{\rb}{\operatorname{rb}}
\newcommand{\R}{\mathbb R}
\newcommand{\C}{\mathbb C}
\newcommand{\CS}{{\mathbb C}S}
\newcommand{\E}{\mathbb E}

\newcommand{\N}{\mathbb N}

\newcommand{\bP}{\mathbb P}

\newcommand{\cA}{\mathcal{A}}

\newcommand{\cH}{\mathcal{H}}

\newcommand{\her}{^{\operatorname h}}
\newcommand{\cM}{\mathcal{M}}

\begin{document}
\bibliographystyle{plain}
\author{Leiba Rodman, Ilya M.\ Spitkovsky,\\ Arleta Szko{\l}a,
Stephan Weis}
\title{Continuity of the maximum-entropy inference:
Convex geometry and numerical ranges approach}
\maketitle
\thispagestyle{empty}
\pagestyle{myheadings}
\markleft{\hfill Continuity of the maximum-entropy inference\hfill}
\markright{\hfill L.~Rodman, I.\,M.~Spitkovsky, A.~Szko{\l}a, S.~Weis \hfill}
\begin{abstract}
We study the continuity of an abstract generalization of the
maximum-entropy inference --- a maximizer. It is defined as a right-inverse
of a linear map restricted to a convex body which uniquely maximizes on
each fiber of the linear map a continuous function on the convex
body. Using convex geometry we prove, amongst others, the existence of
discontinuities of the maximizer at limits of extremal points not being
extremal points themselves and apply the result to quantum correlations.
Further, we use numerical range methods in the case of quantum inference
which refers to two observables. One result is a complete characterization
of points of discontinuity for $3\times 3$ matrices.
\end{abstract}
\vspace{.3in}

\vspace{.2in}

\vspace{.3in}
%
%
%
%
%
%
%
\section{Introduction}
\par
The maximum-entropy principle is a topic in physics since the
19\textsuperscript{th} century in the work of Boltzmann, Gibbs,
von Neumann, Jaynes and many others \cite{vonNeumann,Jaynes}.
The continuity issue of the maximum-entropy
inference under linear constraints in the quantum context was 
studied by one of the authors in \cite{WK} (jointly with A.~Knauf), 
and then further in \cite{Weis-MaxEnt2012}--\cite{Weis-MaxEnt2014}.
Discontinuity points of the maximum-entropy inference are a
distinguished quantum phenomenon because they are
missing in the analogous maximum-entropy inference of probability
distributions which is formally included in the quantum setting
by viewing probability vectors as diagonal matrices.
The discontinuity points were recently discussed by
Chen et al.\ \cite{CJLPSYZZ} in condensed matter physics.
\par
Convex geometry has proved powerful for the problem of discontinuity
in our preceding contribution \cite{Weis-cont}. The crucial difference
with  the simplex of probability distributions is that a quantum state
space, consisting
of density matrices, has curved and flat boundary portions. Already
its planar linear images have a very rich geometry. They correspond to
the notion of {\em numerical range} in operator theory whose shapes
are well understood for $2\times 2$ (the classical elliptical range
theorem) and $3\times 3$ matrices, see the work of Kippenhahn
\cite{Kippenhahn} and earlier papers by the first two authors 
\cite{KRS} (jointly with D.~Keeler) and \cite{RS}.
\par
Let us briefly return to the condensed matter note which is a
substantial motivation for this work.
Definitions of a phase transition in statistical mechanics include
symmetry breaking, such as melting crystals, and long-range
correlations which may be certified by a power law correlation
function, see for example Yeomans \cite{Yeomans1992}. Quantum phase
transitions are not
necessarily associated with symmetry breaking or long-range
correlations, for example see the discussion by Wen \cite{Wen2004},
Sec.~1.4. Still, correlations explain quantum phase
transitions in some cases:
Liu et al.\ \cite{Liu-etal2014} have recognized that a quantum phase
transition in Kitaev's toric code model can be seen in the six-body
correlations of the ground states of a $3\times 4$ torus model.
We demonstrate by an example that our methods are suitable to obtain
analytical results about the correlation quantities used by Liu
et al.\ which are known as {\em irreducible many-party correlations}
and which
were defined by Linden et al.\ and Zhou \cite{Linden-etal2002,Zhou2008}
based on the maximum-entropy principle.
\par
The aim of this article is to contribute to the continuity theory of
the maximum-entropy inference and of the irreducible correlation
using techniques from convex geometry
and operator theory (numerical range). Convex geometrical methods
will be developed in reference to a maximizer $H$ which is defined
as a right-inverse of a linear map $f$ restricted to a convex body
$K$ which uniquely maximizes on each fiber of $f|_K$ a continuous
function on $K$. We have in mind the example where $H$ is an
abstract generalization of the maximum-entropy inference (defined
in Sec.~\ref{sec:preliminaries}) and the continuous function is a
generalization of the von Neumann entropy.
\par
Our analysis will be based on our continuity result \cite{Weis-cont}
which we recall in Sec.~\ref{sec:preliminaries} and which allows us
to study the continuity of the maximizer without solving the respective
inverse problem explicitly. This is possible by studying the
{\em openness} of the restricted linear map $f|_K$. Continuity
results follow as corollaries of openness results.
\par
The results presented in Sec.~\ref{sec:open} have appeared earlier
in the paper \cite{Weis-cont} by the fourth author and are based on 
gauge functions of the domain $L:=f(K)$ of the maximizer. Here we provide
a new unified proof in terms of the notion of {\em simplicial point}
which is a point-wise defined variant of a locally simplicial set in
the sense of Rockafellar \cite{Rockafellar}.
We prove that the maximizer is continuous at all simplicial points ---
in particular it is continuous in the restriction to a polytope or to a
relatively open convex subset.
\par
Further, we present two new results inspired by examples by
Chen et al.\ \cite{CJLPSYZZ}.
In Sec.~\ref{sec:bd-int} we prove a {\em dichotomy} with regard to the
partition of the domain $L$ of the maximizer $H$:
the continuity of $H$ is equivalent to the continuity of its restriction
to the relative boundary of $L$ and can be decided in terms of its
restriction to the relative interior of $L$.
\par
In Sec.~\ref{sec:suff-discont} we prove a necessary condition for the
continuity of $H$ in terms of the {\em face function} studied by Klee and
Martin \cite{KleeMartin1971} and others. The face function maps every
point of $L$ to the unique face of $L$ which contains the given point
in its relative interior. For example, we prove under mild assumptions
(which are satisfied by the maximum-entropy inference) that the maximizer
is discontinuous at points $w\in L$ which are limit points of extremal
points of $L$ but not extremal points themselves. In that case we
remark in Exa.~\ref{ex:non-stable} that the discontinuity at $w$
is not removable from the restriction of $H$ to the relative boundary
of $L$ by changing only the value at $w$.
Discontinuities of the (unrestricted) maximizer $H$ are never
removable as we will point out in Sec.~\ref{sec:bd-int}.
\par
In the three last sections we specialize to the convex body $K$ equal
to a quantum state space.
The aim of Sec.~\ref{sec:irreducible-correlation} is to demonstrate
that convex geometry is an essential part of the topology
of quantum correlations. We use the face function method to point
out discontinuities in the {\em irreducible three-party correlation}
of three qubits which is possible because this correlation quantity
is based on the maximum-entropy principle. We check our analysis by
consulting a result in the context of pure state reconstruction
by Linden et al.\ \cite{Linden-etal2002}.
\par
In Secs.~\ref{sec:singularly-generated}
and~\ref{sec:nr3} we consider two hermitian $d\times d$ matrices,
$d\in\N$, having the meaning of quantum observables. Then the
domain $L$ of the maximizer is planar and can be identified with
the {\em numerical range} of a complex $d\times d$ matrix $A$,
that is $W(A):=\{x^*Ax\mid x\in\CS^d\}$ where $\CS^d$ denotes the
unit-sphere in $\C^d$. We determine the maximum number of
discontinuities in dimension $d=4,5$ (new methods are needed
for $d\geq 6$). Further, in Sec.~\ref{sec:nr3} we give a
complete characterization of points of discontinuity of
the maximum-entropy inference for $d=3$. These results essentially
are corollaries of the Theorem~\ref{thm:md-planar} which connects
to the paper \cite{LLS-OAM} by T.~Leake, B.~Lins and the second 
author, devoted to the numerical range.
One of the proof ideas is that the notion of simplicial point,
mentioned above, is a complement of the notion of
{\em round boundary point} in the second author's papers  
\cite{CJKLS} (jointly with D.~Corey, C.R.~Johnson, 
R.~Kirk and B.~Lins) and \cite{LLS-LMA,LLS-OAM} (jointly
with T.~Leake and B.~Lins) on the numerical range. The second idea 
is concerned with
{\em singularly generated} points which were defined in
\cite{LLS-OAM} as having exactly one linearly independent unit
vector in the pre-image under the map $\CS^d\to\C$, $x\mapsto x^*Ax$.
\par
It is worth mentioning that similar pre-image problems of the numerical range are
of a broader interest, see for example Carden \cite{Carden}. In
particular, pre-image problems for more than two observables appear in
quantum state reconstruction, see for example Gross et al., Heinosaari
et al.\ and Chen et al.\ \cite{Gross-etal2010,Heinosaari-etal2013,CDJJKSZ},
and in quantum chemistry, see for example Erdahl, Ocko et al.\ and
Chen et al.\ \cite{Erdahl1972,Ocko-etal2011,Chen-etal2012}.\\
\par\normalsize
{\par\noindent\footnotesize
{\em Acknowledgements.}
AS and SW appreciate  the support by the German Research Foundation
within the project ``Quantum statistics: decision problems and
entropic functionals on state spaces'' (10/'11--09/'14). IS was
supported in part by the Plumeri Award for Faculty Excellence from
the College of William and Mary and by Faculty Research funding
from the Division of Science and Mathematics, New York University
Abu Dhabi.}
%
%
%
%
\section{Preliminaries}\label{sec:preliminaries}
\setcounter{equation}{0}
\par
In this section we introduce the central objects and concepts of
this article including the notions of state space and maxi\-mum-entropy
inference in quantum mechanics, starting with their abstract counterparts
of convex body and maximizer, as well as the joint numerical range
and the topological notion of open map.
\begin{defn}[Maximizer]\label{def:basic}
Let $X,Y$ be finite-dimensional real norm\-ed vector spaces, let
$f:X\to Y$ be a continuous map and let $K\subset X$ be compact. The
compact set $L:=f(K)\subset Y$ parametrizes the fibers
\[
f|_K^{-1}(w)=\{v\in K\mid f(v)=w\}, \qquad w\in L,
\]
of $f|_K$. We assume the inverse problem of selecting a point in each
fiber is solved by maximizing a continuous function $g:K\to\R$ which
attains a unique maximum on each fiber. The {\em maximizer} is defined
by
\[
H:L\to K,\qquad
w\mapsto{\rm argmax}\{g(v)\mid v\in f|_K^{-1}(w)\}.
\]
All sets which will be introduced in the sequel are
tacitly assumed to be subsets of a finite-dimensional real normed
vector space. Unless otherwise stated we will always assume that $f$
is linear and that $K$ is a {\em convex body}, that is a compact and
convex set.
\end{defn}
\par
We remark that the domain of $f$ is $X$ rather than $K$ by
consistency with our main example (\ref{eq:E}). Since $K$ is
compact, this choice is no restriction because, by the Tietze
extension theorem, any continuous function $K\to Y$ can be
extended to a continuous function $X\to Y$.
\par
On $K\subset X$ and $L=f(K)\subset Y$ we use the subspace topology
induced by $X$ and of $Y$,
respectively.
A {\em neighborhood} of a point $x$ in a topological space $\widetilde{X}$
is any subset of $\widetilde{X}$ containing an open set containing $x$.
We call a map $\gamma:\widetilde{X}\to\widetilde{Y}$ between
topological spaces $\widetilde{X},\widetilde{Y}$ {\em open} at
$x\in\widetilde{X}$ if for any neighborhood $N\subset\widetilde{X}$ of
$x$ the image $\gamma(N)$ is a neighborhood of $\gamma(x)$ in
$\widetilde{Y}$. We say $\gamma$ is {\em open on} a given subset of
$\widetilde{X}$ if $\gamma$ is open at each point in the subset. Finally,
$\gamma$ is {\em open} if $\gamma$ is open on $\widetilde{X}$.
\par
In Thm.~4.9 in \cite{Weis-cont} one of the authors has proved the following.
\begin{fa}[Continuity-openness equivalence]\label{fa:cont-open}
Let $K\subset X$ be an arbitrary compact subset, not necessarily convex,
and let $f:X\to Y$ be an arbitrary continuous function, not necessarily
linear. Then for any $w\in L$ the maximizer $H$ is continuous at $w$ if
and only if $f|_K$ is open at $H(w)$.
\end{fa}
\par
From now on we make the global assumption that $K$ is a convex body and
that $f$ is linear. We will argue in terms of the openness of $f|_K$.
All results may, and some will, be translated into continuity statements
of $H$ using Fact~\ref{fa:cont-open}.
\par
Our main example of convex body $K$ will be the state space of a
matrix algebra which is studied in operator theory \cite{Alfsen-Shultz}.
Let $M_d$, $d\in\N$, denote the full matrix algebra of $d\times d$-matrices
with complex coefficients. The algebra $M_d$ is a complex C*-algebra with
identity $\id_d$. We shall also write $0=0_d$ for the zero in
$M_d$. Let $\cA$ denote a (complex) C*-subalgebra of $M_d$.
For example, we will introduce in Fact~\ref{fa:preimage} the C*-algebras
$pM_dp=\{pap\mid a\in M_d\}$ where $p\in M_d$ is a projection, that is
a hermitian idempotent $p=p^*=p^2$. See Lemma~\ref{lem:cs-small-algebra}
for other relevant examples of C*-algebras.
\par
We denote by $\cA\her=\{a\in\cA\mid a^*=a\}$ the real vector space of
hermitian matrices in $\cA$, known as {\em observables} in physics, and
we endow it with the scalar product
$\langle a,b\rangle:=\trace(ab)$, $a,b\in\cA\her$ which makes
$\cA\her$ a Euclidean space. We call {\em state space} of $\cA$ the
convex body
\[
\cM(\cA):=\{\rho\in\cA\mid\rho\succeq 0,\trace(\rho)=1\}.
\]
Here $a\succeq 0$ means the matrix $a\in\cA$ is positive semi-definite,
that is $a^*=a$ and all eigenvalues of $a$ are non-negative.
Elements of $\cM(\cA)$ are called {\em density matrices} in physics
\cite{Alfsen-Shultz,BZ,NC}. They are in one-to-one correspondence to
the positive normalized linear functionals $\cA\to\C$ called {\em states},
see for example \cite{Alfsen-Shultz}, Sec.~4. We use the terms state and
density matrix synonymously.
\par
Now we confine definitions to the full matrix algebra $M_d$ and we write
$\cM_d:=\cM(M_d)$.
Given a number $r\in\N$ of fixed observables $u_i\in M_d\her$,
$i=1,\ldots,r$, we write ${\bf u}=(u_1,\ldots,u_r)$ and we define
the {\em expected value function}
\begin{equation}\label{eq:E}
\E=\E_{\bf u}:M_d\her\to\R^r,\qquad
a\mapsto (\langle u_1,a\rangle,\ldots,\langle u_r,a\rangle).
\end{equation}
In our earlier papers \cite{Weis-supp,Weis-cont} we have
called the set of expected values
\[
L({\bf u})=L(u_1,\ldots,u_r)
:=\{\E_{\bf u}(\rho)\mid\rho\in\cM_d\}\subset\R^r
\]
the {\em convex support}. This name is motivated by
probability theory \cite{Barndorff}. The probability vectors of length
$d$, embedded as diagonal matrices into $M_d$, are the states of the
algebra of diagonal matrices. Random variables on $\{1,\ldots,d\}$
correspond to diagonal matrices $u_1,\ldots,u_r$ and the set of their
expected value tuples, called convex support in \cite{Barndorff},
equals $L({\bf u})$ by Lemma~\ref{lem:cs-small-algebra}.
\par
Given expected values $\alpha\in L({\bf u})$ the maximum-entropy state
$\rho^*(\alpha)$ is the unique state in the fiber $\E|_{\cM_d}^{-1}(\alpha)$
which maximizes on $\E|_{\cM_d}^{-1}(\alpha)$ the {\em von Neumann entropy}
\[
S(\rho)=-{\rm tr} (\rho \cdot \ln \rho).
\]
Functional calculus with respect to the continuous function $[0,1]\to\R$,
$x\mapsto x\cdot \ln (x)$ where $0 \cdot \ln 0=0$ can be used to define $S$.
The mapping
\begin{equation}\label{eq:maxent}
\rho^*:L({\bf u})\to\cM_d,\qquad
\alpha\mapsto\rho^*(\alpha)
\end{equation}
is called the {\em maximum-entropy inference}, see
\cite{Jaynes,Wichmann,Ingarden} for more details. In physics, the
von Neumann entropy quantifies the uncertainty in a state
\cite{Wehrl1978}. The state $\rho^*(\alpha)$ is considered the most
non-committal, most unbiased or least informative state with
regard to all missing information beyond the expected
values $\alpha$ \cite{Jaynes}.
\par
Although $\rho^*$ can be discontinuous it is smooth up to boundary
points. We denote by $\overline{C}$ the norm closure of any set $C$.
The {\em relative interior} $\ri(C)$ of $C$ is the interior of $C$
in the topology of the affine hull of $C$, and $C$ is
{\em relatively open} if $C=\ri(C)$ holds. The {\em relative boundary}
of $C$ is $\rb(C):=\overline{C}\setminus\ri(C)$. The following
statement is proved in \cite{Wichmann}, Thm.~2b.
\begin{fa}[Real analyticity]\label{1-ref}
The maximum-entropy inference $\rho^*(\alpha)$ is real analytic in the
relative interior of $L({\bf u})$.
\end{fa}
\par
Now we turn to the joint numerical range which will be useful to address
the continuity of $\rho^*$ for two ($r=2$) observables.
It will be convenient to denote the inner product of $x,y\in\C^d$ by
$x^*y:=\overline{x_1}y_1+\cdots+\overline{x_d}y_d$ and to denote by
$xy^*:\C^d\to\C^d$ the linear map defined for $z\in\C^d$ by
$(xy^*)(z):=(y^*z)x$.
Vectors in $\C^n$, $n\in\N$, will be understood as column vectors. To save
space we will write them equivalently in the column and row forms
\[
\left[\begin{array}{c}  x_1 \\x_2 \\ \vdots \\ x_n \end{array}\right]
=(x_1,\ldots,x_n), \qquad
x_1,\ldots,x_n\in\C.
\]
\begin{defn}\label{def:joint-range}
The {\em joint numerical range} of ${\bf u}=(u_1,\ldots, u_r)$ is
the subset of $\R^r$ defined by
\[
W({\bf u})=W(u_1,\ldots,u_r):=
\{(x^*u_1x, \ldots, x^*u_rx)\mid x\in \C^d, x^*x=1\}.
\]
\end{defn}
\par
Let us recall that the convex hull of the joint numerical range
is the convex support. We denote the convex hull of any set $C$
by $\conv(C)$. For all $d,r\in\N$ we
have
\begin{equation}\label{eq:conv-w=l}
\conv(W({\bf u}))=L({\bf u}).
\end{equation}
See \cite{GutkinZyczkowski2013}, Thm.~1, for the identity 
(\ref{eq:conv-w=l}) formulated as an affine isomorphism.
For two observables ($r=2$), if we identify $\R^2\cong\C$, then the joint
numerical range $W(u_1,u_2)$ equals the numerical range $W(u_1+\ii u_2)$
which is convex by the {\em Toeplitz-Hausdorff theorem}. Hence
(\ref{eq:conv-w=l}) implies
\begin{equation}\label{eq:L=W}
L(u_1,u_2)=W(u_1,u_2).
\end{equation}
A proof of (\ref{eq:L=W}) can be found in \cite{BerberianOrland1967}, Thm.~3.
For three observables ($r=3$) the joint numerical range $W(u_1,u_2,u_3)$
is also convex but only for matrix size $d\geq 3$, see \cite{AU,LiPoon}.
\par
Let us discuss easy properties and example of the convex support.
\begin{rem}\label{aug81}
The following transformations do not essentially alter
$L({\bf u})$, $W({\bf u})$, and $\rho^*(\alpha)$.
\begin{enumerate}
\item Remove any $u_i$ which is a (real) linear combination of
$u_1,$ $\ldots,$ $u_{i-1},$ $u_{i+1},$ $\ldots,$ $u_r$.
Conversely, add observables which are linear combinations of
$u_1,\ldots,u_r$.
\item Replace $u_i$ with $u_i+c_i\id$, where $c_i \in \R$, for
any $i$.
\item Replace $u_i$ with $T^*u_iT$ for all $i$, where $T\in M_d$
is a unitary.
\end{enumerate}
\end{rem}
\par
The state space $\cM_2$ of $M_2$ is a three-dimensional
Euclidean ball, known as {\em Bloch ball} \cite{BZ}. Its surface is known
as the {\em Bloch sphere} \cite{BZ,NC} or {\em Poincar\'e sphere}
\cite{Alfsen-Shultz}. The openness of $\E|_{\C\bP^{1}}$ on the Bloch sphere
is proved in Coro.~6 in \cite{CJKLS}. The openness of $\E|_{\cM_2}$ on the
Bloch ball is shown in Example 4.15.2 in \cite{Weis-cont}:
\begin{fa}\label{fa:Bloch}
The expected value function $\E|_{\cM_2}$ is open.
\end{fa}
\par
The possible convex support sets $L({\bf u})$ of $\cM_2$ are the
linear images of the Bloch ball and they are easily identified
algebraically. By Rem.~\ref{aug81}(1,2) we may assume that $r\leq 3$
and that $u_1, u_2, u_3\in M_2\her$ are zero-trace and mutually
orthogonal. If $r=3$, then by Rem.~\ref{aug81}(3) we can take the
observables equal to the {\em Pauli matrices}
\begin{equation}\label{aug82}
\sigma_1:=\left[\begin{array}{cc} 0 & 1 \\ 1 & 0 \end{array}\right], \quad
\sigma_2:=\left[\begin{array}{cc} 0 & -\ii \\ \ii & 0 \end{array}\right], \quad
\sigma_3:=\left[\begin{array}{cc} 1 & 0 \\0 & -1 \end{array}\right].
\end{equation}
It is well known (and easy to check) that $L(\sigma_1,\sigma_2,\sigma_3)$
is the unit ball in $\R^3$. Notice that the joint numerical range
$W(\sigma_1,\sigma_2,\sigma_3)$ is the unit sphere in $\R^3$ which is not
convex. If $r=2$, then we can take $u_1=\sigma_1$, $u_2=\sigma_2$. Then
$L({\bf u})$ is the numerical range of
\[
\sigma_1+\ii\sigma_2 =
\left[\begin{array}{cc} 0 & 2 \\ 0 & 0\end{array}
\right]
\]
which is the unit disk in $\C$ centered at zero. If $r=1$, then $L({\bf u})$
is a line segment (we ignore the trivial case when $L({\bf u})$ is a singleton).
%
%
%
%
\section{Simplicial Points}\label{sec:open}
\setcounter{equation}{0}
\par
In this section we recapitulate results from \cite{Weis-cont} now
with a unified proof in terms of simplicial points. The term of
simplicial point has another advantage that it complements the term
of round boundary point in Lemma~\ref{lem:2d-smoothness}.
\par
The main idea is the gauge condition in Fact~\ref{fa:gauge_condition}
from our work \cite{Weis-cont} which is somewhat similar to the idea
of our Thm.~4 in \cite{CJKLS}. Let $\widetilde{X}$ be a
finite-dimensional real normed vector space and let $C$ be a
convex subset of $\widetilde{X}$. The {\em gauge} of $C$ is defined by
\[
\gamma_C(v):=\inf\{\lambda\geq 0\mid v\in\lambda C\},\quad
v\in\widetilde{X}.
\]
Recall that the gauge of the unit ball in
$\widetilde{X}$ is the norm. More generally, $\gamma_C$ is positively
homogeneous of degree one and convex \cite{Rockafellar}. If
$C\neq\emptyset$ then the {\em positive hull} of $C$ is defined by
${\rm pos}(C):=\{\lambda v\mid\lambda\geq 0, v\in C\}$.
\par
In \cite{Weis-cont}, Prop.~4.11,  one of the authors has proved the 
following.
\begin{fa}[Gauge condition]\label{fa:gauge_condition}
If $w\in L$ and if the gauge $\gamma_{L-w}$ is bounded on the set
of unit vectors in ${\rm pos}(L-w)$ with respect to an arbitrary
norm on $Y$ then $f|_K$ is open on $f|_K^{-1}(w)$.
\end{fa}
\par
The assumptions of Fact~\ref{fa:gauge_condition} are fulfilled for
example at the apices of the skew cone (\ref{eq:papadopoulou}).
The following Prop.~\ref{pro:simplicial} does not apply there but
it will suffice for the purposes of this article.
\par
We call a point $x$ in a convex set $C$ a {\em simplicial point}
if there exists a finite set of simplices $S_1,\ldots,S_m\subset C$
such that the union $S_1\cup\ldots\cup S_m$ is a neighborhood of $x$
in $C$. A convex set $C$ is {\em locally simplicial}
\cite{Rockafellar} if all its elements are simplicial points.
\begin{pro}[Simplicial points]\label{pro:simplicial}
If $w\in L$ is a simplicial point of $L$ then $f|_K$ is open on
$f|_K^{-1}(w)$.
\end{pro}
{\em Proof:}
Let $S_1,\ldots,S_m\subset L$, $m\in\N$, be a set of simplices such
that the union $U:=S_1\cup\ldots\cup S_m$ is a neighborhood of $w$ in
$L$. Since $L$ is convex, the convex hull $P$ of $U$ is also a
neighborhood of $w$ in $L$. Therefore ${\rm pos}(L-w)={\rm pos}(P-w)$
holds. Since $P-w\subset L-w$ holds we have for all vectors
$u\in{\rm pos}(P-w)$ the inequality
\[
\gamma_{L-w}(u)\leq\gamma_{P-w}(u).
\]
For unit vectors $u$ in ${\rm pos}(P-w)$ the right-hand side is bounded
because $P-w$ is polyhedral convex and contains the origin, see
Rem.~3.1 in \cite{Voigt}. Therefore Fact~\ref{fa:gauge_condition}
implies the claim.
\hspace*{\fill}$\square$\\
\par
We mention some examples where we will apply Prop.~\ref{pro:simplicial}.
The relative interior of $L$ and polytopes included in $L$ are locally
simplicial sets \cite{Rockafellar}. So the maximizer $H$ is
continuous on the relative interior of $L$ and globally continuous
if $L$ is a polytope. Moreover, the restriction $H|_P$ is continuous
for every polytope $P\subset L$. For example, the convex support
$L({\bf u})$ is a polytope for commutative observables
$u_1,\ldots,u_r$, see Sec.~2 in \cite{Weis-cont}.
\par
As the last example we mention that relative interior points of facets
are simplicial points and we leave the proof to the reader because
the openness of $f|_K$ on the fibers of these points is also proved in
Coro.~\ref{cor:facet-ri-open} in the next section.

Recall that a {\em face} \cite{Rockafellar} of a
convex set $C$ is any convex subset $F\subset C$ which contains all
segments in $C$ which meet $F$ with an interior point.
A face which is a singleton is called an {\em extremal point}. A
face of codimension one in $C$ is a {\em facet} of $C$.
%
%
%
%
%
\section{Boundary-Interior Dichotomy}\label{sec:bd-int}
\setcounter{equation}{0}
\par
We show that the continuity of the maximizer $H$ is certified by
its restriction to the relative interior $\ri(L)$ of $L$ and also
by the restriction to the relative boundary $\rb(L)$ of $L$.
\par
We begin with the relative interior by citing from Lemma~4.8 in
our paper \cite{Weis-cont}:
\begin{fa}[Norm closure]\label{fa:image-closure}
We have $H(L)\subset\overline{H(\ri(L))}$.
\end{fa}
\par
This statement was proved earlier in the context of the
maximum-entropy inference (\ref{eq:maxent}) in Thm.~2d in
\cite{Wichmann}. The continuity of the maximizer $H$ can be
decided using $H|_{\ri(L)}$. In fact, given $w\in L$, if
\[
f|_K^{-1}(w)\cap \overline{H(\ri(L))}
\]
is a singleton $\{v\}$ then $H$ is continuous at $w$ and $H(w)=v$.
Otherwise $H$ is discontinuous at $w$.
Fact~\ref{fa:image-closure} shows also that discontinuities of $H$
are not removable.
\par
Turning to the relative boundary we will use the property
that for every $w\in\rb(L)$ the convex hull of a neighborhood of
$w$ in $\rb(L)$ and of a point in $\ri(L)$ is a neighborhood of $w$
in $L$. This is easy to check for a Euclidean ball $L$ with center
$w$. The following fact, proved in Sec.~8.1 in \cite{Berge},
generalizes this from the ball to arbitrary convex bodies. Recall
from \cite{Rockafellar} that a mapping $\gamma:\R^n\to\R^n$, $n\in\N$,
is {\em positively homogeneous of degree one} if for each $x\in\R^n$
we have $\gamma(\lambda x)=\lambda\gamma(x)$, $0<\lambda<\infty$.
\begin{fa}[Thm.\ of Sz.~Nagy]%
\label{fa:Nagy}
Let $C\subset\R^n$, $n\in\N$, be a convex body containing the origin
in its interior. Then there exists a homeomorphism from $C$ onto the
standard Euclidean unit ball of $\R^n$ which is positively homogeneous
of degree one.
\end{fa}
\begin{thm}\label{thm:boundary-openness}
Let $\widetilde K:=f|_K^{-1}(\rb(L))$. For all $\widetilde v\in\widetilde K$
the map $f|_K$ is open at $\widetilde v$ if and only if $f|_{\widetilde K}$
is open at $\widetilde v$.
\end{thm}
{\em Proof:}
One direction follows by taking intersections of neighborhoods. Let
us prove conversely that the openness of $f|_{\widetilde K}$ at
$\widetilde v\in{\widetilde K}$ implies the openness of $f|_K$ at
$\widetilde v$.
\par
Let $N\subset K$ be a neighborhood of $\widetilde v$ in $K$. Then
$N\cap\widetilde K$ is a neighborhood of $\widetilde v$ in $\widetilde K$.
By assumptions $f|_{\widetilde K}$ is open at $\widetilde v$ so the image
$f(N\cap\widetilde K)$ is a neighborhood of $f(\widetilde v)$ in $\rb(L)$.
Since $X$ is locally convex, we can assume that $N$ is convex so
$\conv(v,N\cap\widetilde K)\subset N$ for a given point
$v\in N\setminus \widetilde K$. The linearity of $f$ shows
\begin{equation}\label{eq:proof:boundary-continuity}
f(N)\supset f(\conv(v,N\cap\widetilde K))
=\conv(f(v),f(N\cap\widetilde K)).
\end{equation}
Since $f(v)$ lies in the relative interior of $L$ the discussion in the
paragraph before Fact~\ref{fa:Nagy} proves that
$\conv(f(v),f(N\cap\widetilde K))$ is a neighborhood of $f(\widetilde v)$
in $L$. Then (\ref{eq:proof:boundary-continuity}) shows that $f(N)$ is a
neighborhood of $f(\widetilde v)$ in $L$ which completes the proof.
\hspace*{\fill}$\square$\\
\par
Thm.~\ref{thm:boundary-openness} applies to any facet $F$ of $L$
because $F$ is a neighborhood in $\rb(L)$ of the relative
interior points of $F$.
\begin{cor}\label{cor:facet-ri-open}
Let $F$ be a facet of $L$ and let $w\in\ri(F)$. Then $f|_K$ is open on
$f|_K^{-1}(w)$.
\end{cor}
%
%
%
%
%
\section{The Face Function of $L$}\label{sec:suff-discont}
\setcounter{equation}{0}
\par
We prove a necessary continuity condition of the maximizer $H$ in terms
of the lower semi-continuity of the face function of $L$. The lower
semi-continuity of the face function implies that a limit
of extremal points is an extremal point \cite{Papadopoulou1977}.
\par
An example where a limit of extremal points is not an extremal point
is given by the convex hull of
\begin{equation}\label{eq:papadopoulou}
\{(s,t,0)\in\R^3\mid (s-1)^2+t^2=1\}\cup \{(0,0,\pm1)\}.
\end{equation}
Here the set of extremal points
$\{(s,t,0)\in\R^3\mid (s-1)^2+t^2=1, s\neq 0\}$ contains
$(0,0,0)$ which is an interior point of
the segment connecting $(0,0,-1)$ and $(0,0,1)$.
\par
To provide an example with the state space $\cM_3$ we need algebraic
representations of faces.
A subset $F$ of a convex set $C$ is an {\em exposed face} of $C$ if
$F=\emptyset$ or if $F$ equals the set of maximizers in $C$ of a
linear functional. One can show that every exposed face of $C$ is a
face of $C$. An exposed extremal point is called {\em exposed point}.
\begin{fa}[Faces of state spaces]\label{fa:preimage}~
\begin{enumerate}
\item
The non-empty faces of the state space $\cM_d$ are of the form $\cM(pM_dp)$
where $p\in M_d$ is a non-zero projection, see for example \cite{Weis-supp},
Sec.~2.3.
\item
Consider the exposed face $F={\rm argmax}\{\alpha^*\lambda\mid\alpha\in L({\bf u})\}$
of $L({\bf u})$ where $\lambda=(\lambda_1,\ldots,\lambda_r)\in\R^r$. Then
$\E|_{\cM_d}^{-1}(F)=\cM(pM_dp)$ where $p$ is the
spectral projection of the matrix
${\bf u}(\lambda)=\lambda_1u_1+\cdots+\lambda_ru_r$ corresponding to
the maximal eigenvalue of ${\bf u}(\lambda)$. This follows from
$\langle\rho,{\bf u}(\lambda)\rangle=\E(\rho)^*\lambda$, $\rho\in\cM_d$,
$\lambda\in\R^r$, and from \cite{Weis-supp}, Thm.~2.9.
\item
Extremal points of the state space $\cM_d$ are called {\em pure states}.
A state is a pure state if and only if it is of the form $xx^*$ for some
unit vector $x\in\C^d$, see for example \cite{Alfsen-Shultz}, Sec.~4.
\end{enumerate}
\end{fa}
\par
We now discuss a three-dimensional linear image of $\cM_3$ which
has appeared as Exa.~4 in \cite{CJLPSYZZ}. This linear image has
a sequence of extremal points which converge to a point which is
not an extremal point. A three-dimensional cross-section of $\cM_3$
with this property is discussed in Rem~5.9 in \cite{Weis-cont}.
\begin{ex}\label{ex:non-stable}
We consider the convex support $L(u_1,u_2,u_3)$ of
\[
u_1:=\left[\begin{array}{ccc}
1&0&0\\0&1&0\\0&0&-1\end{array}\right],
\quad
u_2:=\left[\begin{array}{ccc}
1&0&1\\0&1&1\\1&1&-1\end{array}\right],
\quad
u_3:=\left[\begin{array}{ccc}
1&0&1\\0&0&1\\1&1&-1\end{array}\right].
\]
\par
{\em A limit of extremal points.}
Let $\epsilon\in\R$ and $\xi(\epsilon):=\sqrt{(1-\epsilon)^2+2\epsilon^2}$.
The eigenvalues of $u_1-\epsilon u_2$ are $\{1-\epsilon,\pm\xi(\epsilon)\}$.
If $\epsilon\neq 0$ then the maximal eigenvalue $\xi(\epsilon)$ is
non-degenerate and $v(\epsilon):=(1,1,\epsilon\cdot x(\epsilon))$ is a
corresponding eigenvector, where
$x(\epsilon):=(\xi(\epsilon)+\epsilon-1)/\epsilon^2$.
By Fact~\ref{fa:preimage}(2) the pure state
$\rho(\epsilon):=c(\epsilon)\cdot v(\epsilon)v(\epsilon)^*$,
where $c(\epsilon)>0$ is for normalization, defines the exposed
point
\[
\alpha(\epsilon):=\E(\rho(\epsilon))
=\tfrac{x(\epsilon)}{2-(1-\epsilon)x(\epsilon)}
(1-\epsilon,
1-3\epsilon,
1-3\epsilon
)-(0,0,\tfrac{1}{2+(\epsilon\cdot x(\epsilon))^2})
\]
of $L({\bf u})$. Since $x(\epsilon)\to1$ for $\epsilon\to 0$ we have
\[\textstyle
\alpha(0):=\lim_{\epsilon\to0}\alpha(\epsilon)=(1,1,\tfrac{1}{2}).
\]
We could also do the easier computation
$\alpha(0)=\E(\lim_{\epsilon\to0}\rho(\epsilon))$ but the focus
should be on $L({\bf u})$ rather than $\cM_3$. The limit $\alpha(0)$ is not
an extremal point because it is
the mid-point of the segment $s\subset L({\bf u})$ between
\[
\E((0,1,0)(0,1,0)^*)=(1,1,0)\quad\mbox{and}\quad
\E((1,0,0)(1,0,0)^*)=(1,1,1).
\]
\par
{\em Discontinuous maximum-entropy inference.}
We would like to point out that the maximum-entropy inference $\rho^*$ is
discontinuous at $\alpha(0)$ along the curve $\alpha(\epsilon)$. Namely,
for $\epsilon\neq 0$ we have
$\rho^*(\alpha(\epsilon))=\rho(\epsilon)$ by Fact~\ref{fa:preimage}(2).
The limit
\[\textstyle
\lim_{\epsilon\to0}\rho^*(\alpha(\epsilon))
=\tfrac{1}2{}(1,1,0)(1,1,0)^*
\]
is a pure state while $\rho^*(\alpha(0))$ has rank two. Indeed, we obtain
\[
\rho^*(\alpha(0))=p/2\quad\mbox{for }p:=(0,1,0)(0,1,0)^*+(1,0,0)(1,0,0)^*.
\]
To see this, observe that the spectral projection corresponding to the maximal
eigenvalue of $u_1$ equals $p$. Hence, by Fact~\ref{fa:preimage}(2),
the exposed face of $L({\bf u})$ consisting of the maximizers of the
linear functional $L({\bf u})\to\R$, $\lambda\mapsto(1,0,0)^*\lambda$
has the pre-image $\cM(pM_3p)$ under $\E|_{\cM_3}$. Since $\alpha(0)=\E(p/2)$
and since $S(p/2)=\log(2)$ is the maximal value of the von Neumann entropy on
$\cM(pM_3p)$ we have $\rho^*(\alpha(0))=p/2$.
\par
{\em Non-removability of the discontinuity.}
Further, we would like to point out that, by Prop.~\ref{pro:simplicial},
the restricted maximum-entropy inference $\rho^*|_s$ is continuous on
the segment $s$ at $\alpha(0)$. This, together with the discontinuity
at $\alpha(0)$ along the curve $\alpha$ proves that the discontinuity
at $\alpha(0)$ is not removable from the restriction of $\rho^*$ to
the relative boundary of $L({\bf u})$ (by changing only the value at
$\alpha(0)$).
\end{ex}

\par
We will show that the discontinuity in Exa.~\ref{ex:non-stable} is
a consequence of the fact that the extremal points $\alpha(\epsilon)$
converge to the point $\alpha(0)$ which lies in the relative interior
of a higher-dimensional face. A convex body $C$ is {\em stable}
\cite{Papadopoulou1977,ClausingPapadopoulou1978} if the mid-point map
\begin{equation}\label{eq:stable}
C\times C\to C, \quad
(x,y)\mapsto\tfrac{1}{2}(x+y)
\end{equation}
is open. The state space $\cM_d$ is stable. Indeed, Lemma~3 in
\cite{Shirokov2006} proves that the map $\cM_d\times\cM_d\times[0,1]$,
$(\rho,\sigma,\lambda)\mapsto (1-\lambda)\rho+\lambda\sigma$ is open.
Prop.~1.1 in \cite{ClausingPapadopoulou1978} then shows that $\cM_d$
is stable.
\par
Let us recall an equivalent statement of stability.
Fact~\ref{fa:face-fn} is proved for example in Thm.~18.2 in
\cite{Rockafellar}.
\begin{fa}[Face function]\label{fa:face-fn}
For each point $x$ in a convex set $C$ there exists a unique face of $C$
which contains $x$ in the relative interior.
\end{fa}
\par
If $x$ is a point in a convex set $C$ then we denote the face of $C$
containing $x$ in its relative interior simply by $F(x)$,
omitting $C$ (which should be clear from the context). The
{\em face function} of $C$ is the set-valued map
\[
C\to C,\quad x\mapsto F(x)
\]
which has been studied for example in
\cite{KleeMartin1971,Papadopoulou1977}. A set-valued map
$\Gamma:\widetilde{X}\to\widetilde{Y}$ between topological spaces
$\widetilde{X},\widetilde{Y}$ is {\em lower semi-continuous} at
$x\in\widetilde{X}$ if for each open set $G$ meeting $\Gamma(x)$
there exists a neighborhood $N$ of $x$ such that for all $x'\in N$
we have $G\cap\Gamma(x')\neq\emptyset$. The set-valued function
$\Gamma$ is {\em lower semi-continuous} if $\Gamma$ is lower
semi-continuous at every point of $\widetilde{X}$.
\begin{fa}[Stable convex bodies]%
\label{fa:lower-semi-face-fn}
If $C$ is a convex body then the following are equivalent
\cite{Papadopoulou1977}:
\begin{enumerate}
\item The convex body $C$ is stable.
\item The face function $x\mapsto F(x)$ of $C$ is lower
semi-continuous.
\item The function $C\to\N_0$, $x\mapsto\dim(F(x))$ is
lower semi-continuous.
\end{enumerate}
\end{fa}
\par
Since extremal points have dimension zero,
Fact~\ref{fa:lower-semi-face-fn}(1) and~(3) prove for any stable
convex body that a limit of extremal points must be an extremal
point. We now list some basic relations between the face functions
of the convex bodies $K$ and $L=f(K)$.
\begin{lem}[Linear images of faces]\label{lem:im-face-fn}
Let $w\in L$. Then:
\begin{enumerate}
\item If $v\in f|_K^{-1}(w)$ then $f(F(v))\subset F(w)$.
\item If $v\in\ri(f|_K^{-1}(w))$ then $f(F(v))=F(w)$.
\end{enumerate}
\end{lem}
{\em Proof:} We prove (1) assuming $v\in f|_K^{-1}(w)$. The point
$v$ is a relative interior point of $F(v)$ by the definition of
the face function. The relative interior of the linear image of
a convex set is the linear image of the relative interior by
\cite{Rockafellar}, Thm.~6.6. So the relative interior of
$f(F(v))$ is $f(\ri(F(v))$. Hence $w=f(v)$ is a relative interior
point of the convex set $f(F(v))$. Therefore, and since $w$ lies in
the face $F(w)$ of $L$, Thm.~18.1 in \cite{Rockafellar} proves (1).
\par
We prove (2) assuming $v\in\ri(f|_K^{-1}(w))$. Recall that inverse
images of faces are faces. So $G:=f|_K^{-1}(F(w))$ is a face of $K$.
As we have recalled in the previous paragraph, $f(\ri(G))=\ri(f(G))$
holds so the affine space $f^{-1}(w)$ meets the relative interior
of $G$. Hence, by \cite{Rockafellar}, Coro.~6.5.1, the relative
interior of $f|_K^{-1}(w)=f^{-1}(w)\cap G$ is the intersection of
$f^{-1}(w)$ with $\ri(G)$. This shows that $v$ lies in $\ri(G)$ and
proves $G=F(v)$ which completes the proof.
\hspace*{\fill}$\square$\\
\par
We are ready for the main result of this section.
\begin{thm}\label{thm:lsc}
Let $(w_i)_{i\in\N}\subset L$ converge to a point $w\in L$. We assume
that {\rm (a)} the convex body $K$ is stable. We also assume that
{\rm (b)} a sequence $(v_i)_{i\in\N}\subset K$ converges to a point
$v\in K$ such that $\E(v_i)=w_i$, $i\in\N$, and such that $v$ lies in
$\ri(f|_K^{-1}(w))$. Then the following statements hold.
\begin{enumerate}
\item
For all open subsets $O\subset L$ meeting $F(w)$ there exists $N\in\N$
such that for all $i\geq N$ we have $O\cap F(w_i)\neq\emptyset$.
\item
We have $\dim F(w)\leq \liminf_{i\to\infty}\dim F(w_i)$.
\end{enumerate}
\end{thm}
{\em Proof:}
We prove (1) assuming $O\subset L$ is an open set meeting $F(w)$.
Since $f|_K$ is continuous $\widetilde{O}:=f|_K^{-1}(O)$ is
open. By the assumption (b) the point $v$ lies in the relative
interior of $f|_K^{-1}(w)$ so $f(F(v))\supset F(w)$ holds by
Lemma~\ref{lem:im-face-fn}(2) and $\widetilde{O}\cap F(v)\neq\emptyset$
follows. By assumption (a) the convex body $K$ is stable so the face
function of $K$ is lower semi-continuous by
Fact~\ref{fa:lower-semi-face-fn}(1) and~(2). As $v=\lim_{i\to\infty}v_i$
holds there is $N\in\N$ such that for $i\geq N$ we have
$\widetilde{O}\cap F(v_i)\neq\emptyset$. As
$f(F(v_i))\subset F(w_i)$ holds by Lemma~\ref{lem:im-face-fn}(1)
we get $O\cap F(w_i)\neq\emptyset$ for $i\geq N$. The statement (2) is
an easy corollary of (1) for arbitrary convex bodies $L$.
\hspace*{\fill}$\square$\\
\par
Let us discuss Thm.~\ref{thm:lsc}.
\begin{rem}\label{rem:discuss-thm}
\begin{enumerate}
\item
Thm.~\ref{thm:lsc} allows us to detect discontinuities of the
maxi\-mum-en\-tro\-py inference $\rho^*$ in terms of the convex
geometry of $L({\bf u})$. We have seen in the paragraph of
(\ref{eq:stable}) that the state space $\cM_d$ is stable while
$\rho^*(\alpha)\in\ri\,\E|_{\cM_d}^{-1}(\alpha)$ holds for all
$\alpha\in L({\bf u})$ by Lemma~5.8 in \cite{Weis-cont}. Thus
the assumptions of Thm.~\ref{thm:lsc} are fulfilled.
The Example~\ref{ex:non-stable} demonstrates explicitly
a discontinuity of $\rho^*$ at a point $\alpha(0)$ which is a
limit of extremal points $\alpha(\epsilon)$ but not an
extremal point itself.
\item
All assumptions of Thm.~\ref{thm:lsc} are needed in general.
Consider a convex body $K\subset X$ which is not stable, take
$X=Y$ and the identity map $f:X\to Y$. Then the face function
of $L=K$ is not lower semi-continuous by
Fact~\ref{fa:lower-semi-face-fn}(1) and (2), thereby
contradicting Thm.~\ref{thm:lsc}(1).
The assumption (b) is not met in Exa.~\ref{ex:non-stable}
where the limit  point
$\rho(0)=\tfrac{1}2{}(1,1,0)(1,1,0)^*$ of
$\rho(\epsilon)\in\cM_3$ for $\epsilon\to0$ is a relative
boundary point of the Bloch ball $\E|_{\cM_3}^{-1}(\alpha(0))$.
In this example we have a jump from the dimension zero of
$F(\alpha(\epsilon))$, $\epsilon\neq 0$, to the dimension one of
$F(\alpha(0))$ which contradicts Thm.~\ref{thm:lsc}(2).
\item
Thm.~\ref{thm:lsc}(1) is in general stronger than Thm.~\ref{thm:lsc}(2)
if the assumptions (a) and (b) of the theorem are ignored. This was
pointed out in the work \cite{Papadopoulou1977} in the example
recalled in (\ref{eq:papadopoulou}) where (1) fails at all points
but the vertices of the segment between $(0,0,-1)$ and $(0,0,1)$
while (2) holds at all points but the mid-point of this segment.
However we do not know whether the convex body in (\ref{eq:papadopoulou})
is the linear image of a stable convex body so (1) and (2) could be
equivalent under the assumptions of the theorem.
\end{enumerate}
\end{rem}
%
%
%
%
\section{Irreducible Correlation}
\label{sec:irreducible-correlation}
\setcounter{equation}{0}
\par
In this section we show that the discontinuity of the three-party
irreducible correlation of three qubits, the correlation that can
not be observed in two-party subsystems \cite{Linden-etal2002},
can be detected {\em via} the face function of the convex body of
two-party marginals.
\par
The discontinuity which we 'detect' follows also from the well-known
result \cite{Linden-etal2002} that almost every pure state of three
qubits is uniquely specified by its two-party
marginals among all states (pure or mixed). The only exceptions
are the GHZ-like states \cite{NC}
\begin{equation}\label{eq:GHZ}
\psi:=\alpha|000\rangle + \beta|111\rangle, \qquad |\alpha|^2+|\beta|^2=1,
\alpha,\beta\in\C
\end{equation}
and their local unitary transforms, that is vectors
$(U_1\otimes U_2\otimes U_3)\psi$ where $U_1,U_2,U_3\in M_2$ are
unitaries.
\par
A three-qubit system ABC is described by the algebra
$M_8\cong M_2\otimes M_2\otimes M_2$ with state space
$\cM_8\cong\cM(M_2\otimes M_2\otimes M_2)$. A {\em two-local Hamiltonian}
is a sum of tensor product terms $a\otimes b\otimes c$ with at most two
non-scalar factors $a,b,c\in M_2\her$. In this section, we fix any
spanning set $u_1,\ldots,u_r\in M_8\her$, $r\in\N$, of the space
$\cH^{(2)}$ of two-local Hamiltonians and put ${\bf u}=(u_1,\ldots,u_r)$.
Given any three-qubit state
$\rho\in\cM_8$, its {\em marginal} $\rho_{AB}\in\cM_4$ on the $AB$
subsystem is defined by
\[
\langle\rho_{AB},a\otimes b\rangle
=\langle\rho,a\otimes b\otimes\id_2\rangle, \qquad
a,b\in M_2.
\]
The marginals $\rho_{AC}$, $\rho_{BC}$ are defined similarly. We denote
by $\rho^{(2)}=(\rho_{AB},\rho_{AC},\rho_{BC})$ the vector of two-party
marginals of $\rho$. For two states $\rho,\sigma\in\cM_8$ the expected
values of two-local Hamiltonians satisfy $\E_{\bf u}(\sigma)=\E_{\bf u}(\rho)$
if and only if the two-party marginals coincide, that is
$\sigma^{(2)}=\rho^{(2)}$. Thus we identify the convex support $L({\bf u})$
with respect to the spanning set $u_1,\ldots,u_r$ of $\cH^{(2)}$ and the
set of two-party marginals, that is
\begin{equation}\label{eq:cs-2body}
L({\bf u})\cong\{\rho^{(2)}\mid\rho\in\cM_8\}.
\end{equation}
Using the identification (\ref{eq:cs-2body}) we define, as in
(\ref{eq:maxent}), the maximum-entropy inference
\[
\rho^*:L({\bf u})\to\cM_8,\qquad
\alpha\mapsto{\rm argmax}\{S(\rho)\in\cM_8\mid\rho^{(2)}=\alpha\}.
\]
The {\em irreducible three-party correlation} \cite{Linden-etal2002} of
$\rho$ is defined as the difference of von Neumann entropies
\begin{equation}\label{eq:irr-corr}
C_3(\rho):=S(\rho^*(\rho^{(2)}))-S(\rho).
\end{equation}
\par
The definition of $C_3(\rho)$ is derived from the statistical inference
view \cite{Jaynes} of the maximum-entropy principle where
$\rho^*(\rho^{(2)})$ is seen as the least informative state compatible
with the two-party marginals $\rho^{(2)}$. Since $\rho$ and
$\rho^*(\rho^{(2)})$ are equal on every two-party subsystem of $ABC$,
any discrepancy between them reveals additional information shared by
$\rho$ which can not be observed on any two-party subsystem. This
information is called {\em irreducible three-party correlation} in
\cite{Linden-etal2002} and is quantified by $C_3(\rho)$.
\par
We remark that information is seen as a constraint on our beliefs
in the context of inference \cite{CatichaGiffin2006} as opposed to
the language usage in coding theory where information is a measure
of unpredictability which is quantified, in the case of quantum
information sources \cite{BjelakovicSzkola2005}, in terms of von
Neumann entropy.
\par
The main point regarding continuity of $C_3$ is provided in
Sec.~5.6 in \cite{Weis-cont}. Lemma~5.15(2) and Lemma~4.5 in
\cite{Weis-cont} prove for all $\alpha\in L({\bf u})$ that
\begin{equation}\label{eq:MaxEnt-IrrCorr}
\begin{array}{c}
\rho^*:L({\bf u})\to\cM_8
\mbox{ is continuous at }\alpha\\
\Updownarrow\\
C_3:\cM_8\to\R
\mbox{ is continuous on }\E|_{\cM_8}^{-1}(\alpha).
\end{array}
\end{equation}
\par
The equivalence (\ref{eq:MaxEnt-IrrCorr}) allows us to apply convex
geometric methods to detect discontinuities of $C_3$. In what follows,
inspired by a model presented in Example~6 in \cite{CJLPSYZZ}, we
prove existence of a discontinuity.
\begin{ex}
We consider the two-local Hamiltonians
\[\begin{array}{c}
H_0:=\id_2\otimes\sigma_3\otimes\sigma_3
+\sigma_3\otimes\id_2\otimes\sigma_3
+\sigma_3\otimes\sigma_3\otimes\id_2\\
H_1:=\sigma_1\otimes\id_2\otimes\id_2
+\id_2\otimes\sigma_1\otimes\id_2
+\id_2\otimes\id_2\otimes\sigma_1
\end{array}\]
where $\sigma_1$ and $\sigma_3$ are Pauli matrices and for $\epsilon>0$
we take
\[
H(\epsilon):=H_0+\epsilon H_1.
\]
The maximal
eigenvalue $\lambda(\epsilon):=1+\epsilon+2\sqrt{1-\epsilon+\epsilon^2}$
of $H(\epsilon)$ is non-degenerate. The positive number
$s=s(\epsilon):=(\lambda(\epsilon)-3)/3\epsilon$ which goes to zero for
$\epsilon\to 0$ allows us to write $w(\epsilon):=(1,s,s,s,s,s,s,1)$ for
the corresponding eigenvector which defines the pure state
$\rho(\epsilon):=w(\epsilon)w(\epsilon)^*/(2+6s^2)$ with $AB$-marginal
\[
\rho(\epsilon)_{AB}
=\tfrac{1}{2}(|00\rangle\langle 00|+|11\rangle\langle 11|)
+\tfrac{1}{2+6s^2}
\left(\begin{smallmatrix}
 -2 s^2 & s^2+s & s^2+s & 2 s \\
 * & 2 s^2 & 2 s^2 & s^2+s \\
 * & * & 2 s^2 & s^2+s \\
 * & * & * & -2 s^2 \\
\end{smallmatrix}\right)
\in\cM_4.
\]
The symmetry of $H(\epsilon)$ shows
$\rho(\epsilon)^{(2)}=(\rho(\epsilon)_{AB},\rho(\epsilon)_{AB},\rho(\epsilon)_{AB})$.
Using the identification (\ref{eq:cs-2body}) we note from
Fact~\ref{fa:preimage}(2) that $\rho(\epsilon)^{(2)}$ is an exposed point
of the convex support $L({\bf u})$ where ${\bf u}=(u_1,\ldots,u_r)$
spans the space of two-local Hamiltonians. The limit $\rho(0)^{(2)}$ of
$\rho(\epsilon)^{(2)}$ for $\epsilon\to0$ is the mid-point of the segment
between the distinct marginals $|000\rangle\langle 000|^{(2)}$ and
$|111\rangle\langle 111|^{(2)}$ and therefore $\rho(0)^{(2)}$ is not an
extremal point of $L({\bf u})$.
\par
Out of curiosity we mention that $\rho(\epsilon)_{AB}$ has rank two
for $0<s<1$
(non-zero eigenvalues $\frac{(s-1)^2}{2 \left(3 s^2+1\right)}$
and $\frac{5 s^2+2 s+1}{2 \left(3 s^2+1\right)}$).
Although the extremal points of $\cM_4$ are rank-one states this does
not contradict the fact that $\rho(\epsilon)^{(2)}$ is an exposed point
of $L({\bf u})$ because $L({\bf u})\subsetneq\cM_4\times\cM_4\times\cM_4$.
\par
Let us turn to the irreducible correlation and its continuity. We have
just seen that $\rho(0)^{(2)}$ is a limit of exposed points but not an
extremal point itself.
Hence Remark~\ref{rem:discuss-thm}(1) shows that the maximum-entropy
inference $\rho^*$ is discontinuous at $\rho(0)^{(2)}$ and
(\ref{eq:MaxEnt-IrrCorr}) proves that the irreducible correlation
$C_3$ is discontinuous at some point in the fiber
$\{\rho\in\cM_8\mid\rho^{(2)}=\rho(0)^{(2)}\}$.
\par
This abstract existence result of a discontinuity is confirmed by
the discontinuity of $C_3$ at the GHZ state
$\tfrac{1}{\sqrt{2}}(|000\rangle + |111\rangle)$ which follows
directly from \cite{Linden-etal2002} and which we explain in detail.
The projection
\[
p:=|000\rangle\langle 000| + |111\rangle\langle 111|
\]
is the spectral projection corresponding to the maximal eigenvalue
of $H_0$. So  Fact~\ref{fa:preimage}(2) applied to
${\bf u}(\lambda)=H_0$ proves that $\cM(pM_8p)$ is the inverse image
of an exposed face of $L({\bf u})$. Elements of the Bloch ball
$\cM(pM_8p)$ can be written in the form
\begin{align*}
\sigma(x,y,z) &:= \frac{1}{2}\big(
p+x\cdot(|000\rangle\langle 111| + |111\rangle\langle 000|)\\
&+y\cdot(-\ii|000\rangle\langle 111| + \ii|111\rangle\langle 000|)\\
&+z\cdot(|000\rangle\langle 000| - |111\rangle\langle 111|)\big)
\end{align*}
where $(x,y,z)\in\R^3$ lies in the three-ball, that is
$x^2+y^2+z^2\leq 1$.  The two-party marginal is
\[
\sigma(x,y,z)^{(2)}=
\tfrac{1}{2}(1+z)(|000\rangle\langle 000|^{(2)}
+\tfrac{1}{2}(1-z)|111\rangle\langle 111|^{(2)}.
\]
The maximum-entropy state
\[
\rho^*(\sigma(x,y,z)^{(2)})=
\tfrac{1}{2}(1+z)(|000\rangle\langle 000|
+\tfrac{1}{2}(1-z)|111\rangle\langle 111|
\]
has von Neumann entropy
$S(\rho^*(\sigma(x,y,z)^{(2)}))=H(\tfrac{1}{2}(1+z))$ where we use
the function
$H(\eta):=-\eta\log(\eta)-(1-\eta)\log(1-\eta)$, $\eta\in[0,1]$.
In particular, for pure states
$\psi:=\alpha|000\rangle+\beta|111\rangle$, $\alpha,\beta\in\C$,
$|\alpha|^2+|\beta|^2=1$ we get
\[
C_3(\psi\psi^*)=H(|\alpha|^2)
\]
which is strictly positive unless $|\alpha|=0$ or $|\alpha|=1$.
The irreducible three-party correlation $C_3(\psi\psi^*)$
has the maximal value $\log(2)$ for states $\psi\psi^*$ where
$\psi=\tfrac{1}{\sqrt{2}}(|000\rangle+e^{\ii\phi}|111\rangle)$,
$\phi\in[0,2\pi)$.
\par
On the other hand, $\psi$ is approximated, for example, by vectors
$\varphi:=\alpha|000\rangle+\cos(\gamma)\beta|111\rangle
+\sin(\gamma)\beta|001\rangle$ for real $\gamma\to 0$. For small
$\gamma>0$ the vector $\varphi$ is not a local unitary transform
of a vector $\psi$ because the two-party marginals of $\varphi\varphi^*$
are not identical. This implies \cite{Linden-etal2002}, as we have
recalled in (\ref{eq:GHZ}), that $\varphi\varphi^*$ is uniquely
determined by its two-party marginals. Hence
$\varphi\varphi^*$ belongs to the maximum-entropy states $\rho^*(L({\bf u}))$
and has zero irreducible three-party correlation. This proves
discontinuity of $C_3$ at $\psi\psi^*$ for all $0<|\alpha|<1$.
Analogous discontinuity statements hold for unitary transforms of
vectors $\psi$.
\end{ex}
%
%
%
%
\section{Multiply Generated Round Boundary Points}
\label{sec:singularly-generated}
\setcounter{equation}{0}
\par
In the sequel we study pairs of observables $u_1,u_2\in M_d\her$,
$d\in\N$, with $\E_{\bf u}:M_d\her\to\R^2$ for ${\bf u}=(u_1,u_2)$
and where $\E_{\bf u}(\cM_d)=L({\bf u})=W({\bf u})$ is the
numerical range (\ref{eq:L=W}). In this section we prove a sufficient
condition for the openness of $\E|_{\cM_d}$ in terms of unique
pre-images and simplicial points. We will see that this condition
works well for matrix sizes $d=3,4,5$ but has a limited meaning
for $d\geq 6$.
\par
In our earlier work \cite{LLS-OAM} we call
$z\in W(u_1+\ii u_2)\subset\C$ {\em singularly generated} if
$x^*(u_1+\ii u_2)x=z$ holds for exactly one linearly independent
unit vector $x\in\C^d$. Otherwise $z$ is {\em multiply generated}.
Since $\E|_{\cM_d}$ is open at those points $\rho\in\cM_d$ where
$\E(\rho)$ is an interior point of the numerical range, a
classification of boundary points is useful. A {\em corner point}
is a point of
$W(u_1,u_2)\subset\R^2$ which belongs to more than one supporting
line of $W(u_1,u_2)$. A {\em flat boundary portion} is a
non-trivial line segment lying in the boundary of $W(u_1,u_2)$.
A boundary point of $W(u_1,u_2)$ which is no corner point and
which does not belong to the relative interior of a flat boundary
portion is called {\em round boundary point}.
\par
Notice that every flat boundary portion is a sub-segment of a
one-dimensional face of $W(u_1,u_2)$. Round boundary points exist
only if the numerical range has dimension two.
\begin{lem}\label{lem:2d-smoothness}
If the dimension of $W(u_1,u_2)$ is two then every corner point is
the intersection of two facets of $W(u_1,u_2)$. Without dimension
restrictions, every point of $W(u_1,u_2)$ is either a round boundary
point or a simplicial point but not both.
\end{lem}
{\em Proof:}
Every corner point $\alpha\in W(u_1+\ii u_2)$ is a
{\em normal splitting eigenvalue},
see Sec.~13 in \cite{Kippenhahn}, that is $u_1+\ii u_2$ is unitarily
equivalent to a block diagonal matrix
\[
\left[\begin{array}{c|c}
\alpha & \\\hline
 & B
\end{array}\right]
\]
with zeros on the off-diagonal. Since $W(u_1+\ii u_2)$ is the convex
hull of $W(B)$ and of $\alpha$ the first statement follows by induction.
Going through the above classification of boundary points, the second
assertion now follows easily.
\hspace*{\fill}$\square$\\
\begin{rem}[Lattice theoretical proof of Lemma~\ref{lem:2d-smoothness}]
The numerical range $W(u_1,u_2)$ is the convex dual of an affine
section of $\cM_d$ \cite{Henrion2010,Weis-supp,HeltonSpitkovsky2012}.
Since all faces of this affine section are exposed, the lattice
isomorphism (2) in \cite{Weis-touch} shows that all non-empty faces of
normal cones of $W(u_1,u_2)$ are normal cones. In particular, if
$\alpha$ is a corner point then the two boundary rays of its normal
cone are normal cones of $W(u_1,u_2)$. The lattice
isomorphism (1) in \cite{Weis-touch} now shows that $\alpha$ is
an extremal point of two distinct one-dimensional faces of
$W(u_1,u_2)$ which proves the claim.
\end{rem}
\begin{thm}\label{thm:md-planar}
The map $\E_{\bf u}|_{\cM_d}$ is open on $\cM_d$ except possibly at
those states $\rho\in\cM_d$ where $\E_{\bf u}(\rho)$ is a multiply
generated round boundary point.
\end{thm}
{\em Proof:}
If $\alpha\in W(u_1,u_2)$ is singularly generated then by Thm.~5 in
\cite{CDJJKSZ} the fiber $\E|_{\cM_d}^{-1}(\alpha)$ is a singleton.
It is easy to prove and well-known in the theory of multi-valued maps
\cite{CJKLS} that $\E|_{\cM_d}$ is open at singleton fibers. If
$\alpha$ is a simplicial point then Prop.~\ref{pro:simplicial} proves
that $\E|_{\cM_d}$ is open on $\E|_{\cM_d}^{-1}(\alpha)$. Otherwise,
if $\alpha$ is not a simplicial point, then the second statement of
Lemma~\ref{lem:2d-smoothness} shows that $\alpha$ is a round
boundary point.
\hspace*{\fill}$\square$\\
\par
The converse of Thm.~\ref{thm:md-planar} does not hold for $d\geq 4$,
that is, $\E|_{\cM_d}$ may be open on fibers of multiply generated
round boundary points.
\begin{ex}
An example inspired by Thm.\ 4.4 in
\cite{LLS-OAM} is
\begin{equation}\label{eq:4x4double-ellipse}
u_1+\ii u_2
=\left[\begin{array}{cc} 0 & 2 \\ 0 & 0 \end{array}\right]
\oplus
\left[\begin{array}{cc} 0 & 2 \\ 0 & 0 \end{array}\right]
=\id_2\otimes
\left[\begin{array}{cc} 0 & 2 \\ 0 & 0 \end{array}\right],
\end{equation}
where the numerical range is the unit disk and all points on the unit
circle $S^1$ are multiply generated round boundary points. Nevertheless
$\E|_{\cM_4}:\cM_4\to\R^2$ is open. A short computation with the unitary
\[
v_\theta:=\id_2\otimes[\cos(\tfrac{\theta}{2})\id_2
-\ii \sin(\tfrac{\theta}{2})\sigma_3]\in M_2,
\qquad \theta\in\R,
\]
and the Pauli matrix $\sigma_3$ shows
\[
\E(v_\theta\rho v_\theta^*)=\left[\begin{array}{cc} \cos(\theta) & -\sin(\theta) \\
\sin(\theta) & \cos(\theta) \end{array}\right] \cdot \E(\rho),
\qquad \rho\in\cM_4.
\]
The state space $\cM_4$ is partitioned into the orbits of the group of
unitaries $\{v_\theta\mid\theta\in[0,2\pi)\}$. The orbit $\mathcal O(\rho)$
through any state $\rho$ with $\E(\rho)\in S^1$ is homeomorphic to $S^1$
under $\E$. Hence $\E|_{\mathcal O(\rho)}$ is open at $\rho$ and
{\em a fortiori} $\E$ restricted to $\E|_{\cM_4}^{-1}(S^1)$ is open at
$\rho$. Now Thm.~\ref{thm:boundary-openness} proves that $\E|_{\cM_4}$
is open at $\rho$. The openness of $\E|_{\cM_4}$ at all other states
of $\cM_4$ follows from Prop.~\ref{pro:simplicial} applied to the
interior of the unit disk.
\end{ex}
\par
To capture the non-generic direct sum form (\ref{eq:4x4double-ellipse}) we
introduce a definition from \cite{KRS}.
A matrix $A\in M_d$ is {\em unitarily reducible} if $A$ is unitarily
equivalent to a matrix in block-diagonal form with at least two
proper blocks. Otherwise $A$ is {\em unitarily irreducible}.
\par
Unitarily irreducible matrices have at most $d-3$ multiply-generated
round boundary points when $d=3,4,5$. See Thm.~3.2 in \cite{LLS-OAM}
for $d=3$ and notice that every round boundary point is an extremal
point. For $d=4,5$ see Thms.~4.1 and~5.7 in \cite{LLS-OAM}.
Since $\E|_{\cM_2}$ is open by Fact~\ref{fa:Bloch} and since $\cM_1$
is a singleton, Thm.~\ref{thm:md-planar} implies the following.
\begin{cor}\label{cor:ub}
Let $d\leq 5$ and let $u_1+\ii u_2$ be unitarily irreducible. Then there
are at most $\max\{0,d-3\}$ points $\alpha$ of $W(u_1,u_2)$ such that
$\E|_{\cM_d}$ is not open on the fiber $\E|_{\cM_d}^{-1}(\alpha)$.
\end{cor}
\par
The maximum-entropy inference $\rho^*:W(u_1,u_2)\to\cM_d$ is indeed
discontinuous at the multiply generated round boundary point(s) of
$W(u_1,u_2)$, if there are any, when $d=4$ or $5$ and when $u_1+\ii u_2$
is unitarily irreducible. This happens because these points are isolated.
So, approximating a multiply generated round boundary point
$\alpha$ by singularly generated extremal points $\alpha_i$ we observe
that $\rho^*(\alpha_i)$ has rank one while the rank of $\rho^*(\alpha)$
is at least two. This proves discontinuity of $\rho^*$ at $\alpha$.
\par
It is known for matrix size $d\geq 6$ and irreducible matrix $u_1+\ii u_2$
that $W(u_1,u_2)$ may have infinitely many multiply generated round
boundary points. As was noticed earlier in \cite{LLS-OAM}, Sec.~6,
this is closely connected to the failure of {\em Kippenhahn's conjecture}.
On the other hand, one of us has recently shown \cite{Weis2015} that 
for all $d\in\N$ the maximum-entropy inference $\rho^*:W(u_1,u_2)\to\cM_d$ 
has at most finitely many discontinuities.
%
%
%
%
\section{Numerical Range of $3 \times 3$ Matrices}\label{sec:nr3}
\setcounter{equation}{0}
\par
In the numerical range approach, based on \cite{Kippenhahn,KRS,RS}, we
characterize points of discontinuity of  $\rho^*(\alpha)$ depending on
the type of shape of the numerical range of $3 \times 3$ matrices:
unitarily irreducible --- ovular, ellipse,
with a flat portion on the boundary, and unitarily reducible --- triangle,
line segment, ellipse, and the convex hull of an ellipse and a point
outside the ellipse.
\par
See Example 4.18 in \cite{Weis-cont} for the problem of openness of
$\E|_{\cM(\cA)}$ for unitarily reducible $u_1+\ii u_2$ and the C*-algebra
$\cA\subset M_3$ generated by $u_1+\ii u_2$. Here we consider the
full algebra $M_3$ and arbitrary $3\times 3$ matrices.
\begin{thm}\label{thm:3x3}
Let $u_1,u_2\in M_3\her$. If $\E|_{\cM_3}$ is not open on the fiber
$\E|_{\cM_3}^{-1}(z)$ of a point $z\in W(u_1+\ii u_2)$ then after
reparametrization {\rm (\ref{aug81})} the matrix $u_1+\ii u_2$ has
the form
\[
A:=\left[\begin{array}{ccc} 0 & 2 \\ 0 & 0 \end{array}
\right]\oplus [\,1\,]
\]
and $W(A)$ is the unit disk in $\C$. The map $\E|_{\cM_3}$ is open on
$\cM_3$ with the exception of the fiber $\E|_{\cM_3}^{-1}(1)$ which
is a three-dimensional ball where $\E|_{\cM_3}$ is only open at the
pure state $v_1v_1^*$ for $v_1:=\tfrac{1}{\sqrt{2}}(1,1,0)$.
\end{thm}
{\em Proof.}
If $\E|_{\cM_3}$ is not open on $\E|_{\cM_3}^{-1}(z)$ then
Thm.~\ref{thm:md-planar} shows that $z$ is a multiply generated
round boundary point. Thm.~3.2 in \cite{LLS-OAM} shows then that
$u_1+\ii u_2$ is unitarily equivalent to a matrix $B\oplus [z]$
where $B$ is a
unitarily irreducible $2\times 2$ matrix and $z$ is a boundary
point of the ellipse $W(u_1+\ii u_2)$. This and (\ref{aug81})
allow us to transform the matrix $u_1+\ii u_2$ into the above
matrix $A$. Now $W(u_1+\ii u_2)$ is the unit
disk and $1$ is a multiply generated round boundary point. By
Thm.~3.2 in \cite{LLS-OAM} all points $\alpha\neq 1$ on the unit
circle $S^1$ are singularly generated, so the fiber
$\E|_{\cM_3}^{-1}(\alpha)$ is a singleton by Thm.~5 in
\cite{CDJJKSZ}. At singleton fibers $\E|_{\cM_3}$ is open. The
map $\E|_{\cM_3}$ is open on the fibers of all interior points
of the unit disk, see Prop.~\ref{pro:simplicial}, so it remains
to examine the exceptional point $1\in S^1$.
\par
Since for any $\alpha\in S^1\setminus\{1\}$ the fiber of $\alpha$
is a singleton we have $H(\alpha)=v_\alpha v_\alpha^*$ where
$v_\alpha:=\tfrac{1}{\sqrt{2}}(1,\alpha,0)$.
Choosing any $\rho_0\in\E|_{\cM_3}^{-1}(1)$ and maximizing the
quadratic form
$g_{\rho_0}(\rho):=-\langle \rho-\rho_0,\rho-\rho_0\rangle$,
$\rho\in\cM_3$, gives $H(1)=\rho_0$. The restriction $H|_{S^1}$
is continuous at $1$ if and only if $\rho_0=v_1v_1^*$. This, by
Thm.~\ref{thm:boundary-openness}, is also the condition that
$H$ is continuous at $1$ and, by Fact~\ref{fa:cont-open}, the
condition that $\E|_{\cM_3}$ is open at $\rho_0$.
\hspace*{\fill}$\square$\\
\par
The maximum-entropy states with respect to the transformed
observables $u_1=\sigma_1\oplus 1$ and $u_2=\sigma_2\oplus 0$
in Thm.~\ref{thm:3x3} are
\[
\rho^*(1)=\left[\begin{array}{cc} 1/4 & 1/4 \\
1/4 & 1/4 \end{array}\right] \oplus [\,1/2\,]
\]
and
\[
\rho^*(\alpha)=\left[\begin{array}{cc} 1/2 & \overline{\alpha}\cdot 1/2  \\
\alpha\cdot 1/2 & 1/2 \end{array}\right]\oplus [\,0\,],
\qquad |\alpha|=1, \quad \alpha \neq 1.
\]
This proves discontinuity of $\rho^*$ at $\alpha=1$.
\par
We conclude with a remark about several observables $u_1,\ldots,u_r\in M_3\her$,
$r\in\N$, and ${\bf u}=(u_1,\ldots,u_r)$.
\begin{pro}\label{pro:facet-bd}
Let $F$ be a face of $L({\bf u})=\E(\cM_3)$ and assume
$0<\dim(F)<\dim(L({\bf u}))$. Then $F$ is a Euclidean ball of dimension
one, two or three, and all relative boundary points $w\in\rb(F)$ have
singleton fibers $\E|_{\cM_3}^{-1}(w)$.
\end{pro}
{\em Proof:} Let $w$ be an extremal point of $F$. Then
$\{w\}\subset F\subset L$ are faces of $L$, properly included
into each other. Since inverse images of faces are faces,
Fact~\ref{fa:preimage}(1) shows
\[
\E|_{\cM_3}^{-1}(F)=\cM(p_1M_3p_1),\qquad
\E|_{\cM_3}^{-1}(\{w\})=\cM(p_2M_3p_2)
\]
for projections $p_1\succeq p_2$. By the assumption of strict
dimension differences we get
\[
3=\rank(\id_3)>\rank(p_1)>\rank(p_2)\geq 1.
\]
Thus
$\rank(p_1)=2$ holds and so $\E|_{\cM_3}^{-1}(F)$ is a copy
of the three-dimensional Euclidean Bloch ball. Now the
claim follows easily.
\hspace*{\fill}$\square$\\
\par
Prop.~\ref{pro:facet-bd} implies the following (for
several observables, $r\in\N$).
\begin{cor}\label{cor:3x3facet-open}
Let $F$ be a facet of $L({\bf u)}=\E(\cM_3)$. Then $\E|_{\cM_3}$
is open on $\E|_{\cM_3}^{-1}(F)$.
\end{cor}
\par
An example where Coro.~\ref{cor:3x3facet-open} gives new insights
beyond Coro.~\ref{cor:facet-ri-open} is the convex hull of the
Steiner Roman surface, depicted in Fig.~10 in \cite{BWZ}, which is
a linear image of $\cM_3$ and which has four disk facets.
%
%
%
%
\section{Appendix}\label{sec:appendix}
\setcounter{equation}{0}
\par
We recall a reduction of the state space in terms of the algebra
of observables. The proof is from Sec.~3.4 in \cite{Weis-supp}
and is reproduced here in a simpler setting.
\par
For $d\in\N$ let us consider the parametrization of the hermitian pencil
\[
{\bf u}(\theta):=\theta_1u_1+\cdots+\theta_ru_r,
\qquad \theta\in\R^r
\]
for observables $u_1,\ldots,u_r\in M_d\her$. Let $\cA({\bf u})$ denote the
real or complex *-algebra generated by the $d\times d$ identity matrix
$\id_d$ and by $u_1,\ldots,u_r$. Recall that Minkowski's theorem asserts
that a convex body is the convex hull of its extremal points.
The statement of Straszewicz's theorem is that the closure of exposed points
of a convex body contains all its extremal points. See e.g.\
Coro.~1.4.5 and Thm.~1.4.7 in \cite{Schneider} for these statements.
\begin{lem}\label{lem:cs-small-algebra}
Any spectral projection of any matrix in the hermitian pencil
$\{{\bf u}(\theta)\mid\theta\in\R^r\}$ belongs to $\cA({\bf u})$.
The convex support of ${\bf u}$ is $L({\bf u})=\E(\cM(\cA({\bf u})))$.
\end{lem}
{\em Proof:}
Any spectral projection of ${\bf u}(\theta)$ for $\theta\in\R^r$ is a real
polynomial in one variable evaluated at ${\bf u}(\theta)$. Therefore the
spectral projection belongs to the algebra $\cA({\bf u})$.
\par
Let $\alpha$ be an exposed point of $L({\bf u})$. Then
Fact~\ref{fa:preimage}(2) shows that there is $\theta\in\R^r$ such that the
spectral projection $p$ of ${\bf u}(\theta)$, corresponding to the largest
eigenvalue of of ${\bf u}(\theta)$, yields
\[
\E|_{\cM_d}^{-1}(\alpha)=\cM(pM_dp).
\]
Since $p/\trace(p)$ belongs to the algebra $\cA:=\cA({\bf u})$,
the image of the state space $\cM(\cA)$ under $\E$ covers all
exposed points of $L({\bf u})$. Now Straszewicz's theorem implies
that $\cM(\cA)$ covers all extremal points of $L({\bf u})$, and
thus by Minkowski's theorem the whole $L({\bf u})$. The converse
inclusion is trivial because $\cA\subset M_d$ holds.
\hspace*{\fill}$\square$\\
%
%
%


\begin{thebibliography}{10}

\bibitem{Alfsen-Shultz} E.\,M.~Alfsen, F.\,W.~Shultz (2001)
{\em State Spaces of Operator Algebras:\ Basic Theory, Orientations,
and C*-Products}, Springer-Verlag

\bibitem{AU} Y.\,H.~Au-Yeung, Y.\,T.~Poon (1979)
{\em A remark on the convexity and positive definiteness
concerning Hermitian matrices},
Southeast Asian Bull. Math. 3 85--92

\bibitem{Barndorff} O.~Barndorff-Nielsen (1978)
{\em Information and Exponential Families in Statistical Theory},
John Wiley \& Sons, New York

\bibitem{BWZ} I.~Bengtsson, S.~Weis, K.~\.Zyczkowski (2013)
{\em Geometry of the set of mixed quantum states:\ An apophatic approach},
in Geometric Methods in Physics, Basel 175--197

\bibitem{BZ} I.~Bengtsson, K.~\.Zyczkowski (2006)
{\em Geometry of Quantum States}, Cambridge University Press

\bibitem{BerberianOrland1967} S.\,K.~Berberian, G.\,H.~Orland (1967)
{\em On the closure of the numerical range of an operator},
Proc Amer Math Soc 18(3) 499-503

\bibitem{Berge} C.~Berge (1963)
{\em Topological Spaces},
Edinburgh-London:\ Oliver \& Boyd

\bibitem{BjelakovicSzkola2005} I.~Bjelakovi\'c, A.~Szko{\l}a (2005)
{\em The data compression theorem for ergodic quantum information sources},
Quantum Information Processing 4(1) 49--63

\bibitem{Carden} R.~Carden (2009)
{\em A simple algorithm for the inverse field of values problem},
Inverse Problems 25(11) 115019

\bibitem{CatichaGiffin2006} A.~Caticha, A.~Giffin (2006)
{\em Updating Probabilities},
AIP Conf.\ Proc.\ 872 31--42

\bibitem{Chen-etal2012} J.\ Chen, Z.\ Ji, M.\,B.\ Ruskai, B.\ Zeng,
D.-L.\ Zhou (2012)
{\em Comment on some results of Erdahl and the convex structure of
reduced density matrices},
Journal of Mathematical Physics 53(7) 072203

\bibitem{CDJJKSZ} J.~Chen, H.~Dawkins, N.~Johnston, D.~Kribs,
F.~Shultz, B.~Zeng (2013)
{\em Uniqueness of quantum states compatible with given measurement results},
Phys Rev A 88(1) 012109

\bibitem{CJLPSYZZ} J.~Chen, Z.~Ji, C.-K.~Li, Y.-T.~Poon, Y.~Shen,
N.~Yu, B.~Zeng, D.~Zhou (2014)
{\em Principle of maximum entropy and quantum phase transitions},
\verb+arXiv:1406.5046+[quant-ph]

\bibitem{ClausingPapadopoulou1978} A.~Clausing, S.~Papadopoulou (1978)
{\em Stable convex sets and extremal operators},
Mathematische Annalen 231 193--203

\bibitem{CJKLS} D.~Corey, C.\,R.~Johnson, R.~Kirk, B.~Lins,
I.~Spitkovsky (2013)
{\em Continuity properties of vectors realizing points in the classical
field of values},
Linear and Multilinear Algebra 61(10) 1329--1338

\bibitem{Erdahl1972} R.\,M.\ Erdahl (1972)
{\em The convex structure of the set of N-representable reduced 2-matrices},
Journal of Mathematical Physics 13(10) 1608--1621

\bibitem{Gross-etal2010} D.\ Gross, Y.-K.\ Liu, S.\,T.\ Flammia,
S.\ Becker, J.\ Eisert (2010)
{\em Quantum state tomography via compressed sensing},
Phys Rev Lett 105(15) 150401

\bibitem{GutkinZyczkowski2013} E.\ Gutkin, K.\ \.Zyczkowski (2013)
{\em Joint numerical ranges, quantum maps, and joint numerical shadows},
Linear Algebra and its Applications 438(5) 2394--2404

\bibitem{Heinosaari-etal2013} T.\ Heinosaari, L.\ Mazzarella,
M.\,M.\ Wolf (2013)
{\em Quantum tomography under prior information},
Communications in Mathematical Physics 318(2)
355--374

\bibitem{HeltonSpitkovsky2012} J.\,W.~Helton, I.\,M.~Spitkovsky
(2012) {\em The possible shapes of numerical ranges},
Operators and Matrices 6(3) 607--611

\bibitem{Henrion2010} D.~Henrion (2010)
{\em Semidefinite geometry of the numerical range},
Electronic Journal of Linear Algebra 20 322--332

\bibitem{Ingarden} R.\,S.~Ingarden, A.~Kossakowski, M.~Ohya (1997)
{\em Information Dynamics and Open Systems}, Kluwer Academic Publishers Group

\bibitem{Jaynes} E.\,T.~Jaynes (1957)
{\em Information theory and statistical mechanics.},
Phys Rev 106 620--630 and 108 171--190

\bibitem{Kippenhahn} R.~Kippenhahn (1951)
{\em \"Uber den Wertevorrat einer Matrix},
Math Nachr 6 193--228

\bibitem{KleeMartin1971} V.~Klee, M.~Martin (1971)
{\em Semicontinuity of the face-function of a convex set},
Comm Math Helv 46(1) 1--12

\bibitem{KRS} D.\,S.~Keeler, L.~Rodman, I.\,M.~Spitkovsky (1997)
{\em The numerical range of $3\times 3$ matrices},
Lin Alg Appl 252 115--139

\bibitem{LLS-LMA} T.~Leake, B.~Lins, I.\,M.~Spitkovsky (2014)
{\em Inverse continuity on the boundary of the numerical range},
Linear and Multilinear Algebra 62 1335--1345

\bibitem{LLS-OAM} T.~Leake, B.~Lins, I.\,M.~Spitkovsky (2014)
{\em Pre-images of boundary points of the numerical range},
Operators and Matrices 8 699--724

\bibitem{LiPoon} C.-K.~Li, Y.-T.~Poon (2000)
{\em Convexity of the joint numerical range},
SIAM J Matrix Anal A 21(2) 668--678

\bibitem{Linden-etal2002} N.~Linden, S.~Popescu, W.~Wootters (2002)
{\em Almost every pure state of three qubits is completely determined
by its two-particle reduced density matrices},
Phys Rev Lett 89(20) 207901

\bibitem{Liu-etal2014} Y.\ Liu, B.\ Zeng, D.\,L.\ Zhou (2014)
{\em Irreducible many-body correlations in topologically ordered systems},
\verb+arXiv:1402.4245+[quant-ph]

\bibitem{NC} M.\,A.\ Nielsen, I.\,L.\ Chuang (2000)
{\em Quantum Computation and Quantum Information},
Cambridge University Press

\bibitem{Ocko-etal2011} S.\,A.\ Ocko, X.\ Chen, B.\ Zeng, B.\ Yoshida,
Z.\ Ji, M.\,B.\ Ruskai, I.\,L.\ Chuang (2011)
{\em Quantum codes give counterexamples to the unique preimage conjecture
of the N-representability problem},
Phys Rev Lett 106(11) 110501

\bibitem{Papadopoulou1977} S.~Papadopoulou (1977)
{\em On the geometry of stable compact convex sets},
Math Ann 229 193--200

\bibitem{Rockafellar} R.\,T.~Rockafellar (1972)
{\em Convex Analysis},
Princeton University Press

\bibitem{RS} L.~Rodman, I.\,M.~Spitkovsky (2005)
{\em $3\times 3$ matrices with a flat portion on the boundary of the
numerical range}, Lin Alg Appl 397 193--207

\bibitem{Schneider} R.~Schneider (2014)
{\em Convex Bodies:\ The Brunn-Minkowski Theory}, 2nd Edition,
Cambridge University Press

\bibitem{Shirokov2006} M.\,E.~Shirokov (2006)
{\em The Holevo capacity of infinite dimensional channels and the
additivity problem}, Commun Math Phys 262 137--159

\bibitem{Voigt} I.~Voigt, S.~Weis (2010)
{\em Polyhedral Voronoi cells},
Contrib.\ Algebra and Geometry 51, 587--598

\bibitem{vonNeumann} J.~von~Neumann (1927)
{\em Thermodynamik quantenmechanischer Gesamtheiten},
G\"ott Nach 273--291

\bibitem{Wehrl1978} A.~Wehrl (1978)
{\em General properties of entropy},
Rev Modern Phys 50(2) 221--260

\bibitem{Weis-supp} S.~Weis (2011)
{\em Quantum convex support},
Lin Alg Appl 435 3168--3188;
correction (2012) ibid.\ 436 xvi

\bibitem{Weis-touch} S.~Weis, (2012)
{\em A note on touching cones and faces},
Journal of Convex Analysis 19(2) 323--353

\bibitem{Weis-MaxEnt2012} S.~Weis (2013)
{\em Discontinuities in the maximum-entropy inference},
AIP Conference Proceedings 1553 192--199

\bibitem{Weis-cont} S.~Weis (2014)
{\em Continuity of the maximum-entropy inference},
Communications in Mathematical Physics 330(3) 1263--1292

\bibitem{Weis-MaxEnt2014} S.~Weis (2015)
{\em The MaxEnt extension of a quantum Gibbs family,
convex geometry and geodesics},
AIP Conference Proceedings 1641 173--180

\bibitem{Weis2015} S.~Weis (in preparation)
{\em Maximum-entropy inference and
inverse continuity of the numerical range}

\bibitem{WK} S.~Weis, A.~Knauf (2012)
{\em Entropy distance:\ New quantum phenomena},
J Math Phys 53(10) 102206

\bibitem{Wen2004} X.-G.~Wen (2004)
{\em Quantum Field Theory of Many-Body Systems},
Oxford University Press (2004)

\bibitem{Wichmann} E.\,H.~Wichmann (1963)
{\em Density matrices arising from incomplete measurements},
J Math Phys 4(7) 884--896

\bibitem{Yeomans1992} J.\,M.\ Yeomans (1992)
{\em Statistical Mechanics of Phase Transistions},
Oxford University Press Inc., New York

\bibitem{Zhou2008} D.~Zhou (2008)
{\em Irreducible multiparty correlations in quantum states without
maximal rank}, Physical Review Letters 101 180505

\end{thebibliography}

\vspace{.5in}

\begin{center}
\begin{tabular}{l}
Leiba Rodman \\ 
Department of Mathematics \\
College of William and Mary \\
P. O. Box 8795 \\
Williamsburg, VA 23187-8795 \\
e-mail: lxrodm@math.wm.edu, lxrodm@gmail.com \\
~\\
Ilya M.\ Spitkovsky \\
Division of Science and Mathematics \\
New York University Abu Dhabi \\
Saadiyat Island, P.O. Box 129188 \\
Abu Dhabi, UAE \\
e-mail: ims2@nyu.edu \\
~\\
and\\
~\\
Department of Mathematics \\
College of William and Mary \\
P.\, O.~Box 8795 \\
Williamsburg, VA 23187-8795 \\
e-mail: ilya@math.wm.edu \\
~\\
Arleta Szko{\l}a \\
Max-Planck-Institute for \\
Mathematics in the Sciences \\
Inselstrasse 22 \\
D-04103 Leipzig \\
Germany \\
e-mail: szkola@mis.mpg.de \\
~\\
Stephan Weis \\
Inselstrasse 28 \\
D-04103 Leipzig \\
Germany \\
e-mail: maths@stephan-weis.info
\end{tabular}
\end{center}
\end{document}